\documentstyle[astrobib]{mnv2}

\input epsf
\input rotate
\epsfverbosetrue

\newcommand{\sigmaT}{\sigma_{\rm {\scriptscriptstyle T}}}
\newcommand{\lambdaT}{\lambda_{\rm {\scriptscriptstyle T}}}
\newcommand{\vT}{v_{\rm {\scriptscriptstyle T}}}

\title[Constraints on Clouds in AGN]
      {Physical constraints on the sizes of dense clouds
       in the central magnetospheres of Active Galactic Nuclei}
\author[Kuncic, Blackman and Rees]
       {Z. Kuncic\thanks{Present address: Astrophysical Theory Centre,
        School of Mathematical Sciences, Australian National University,
        Canberra, ACT 0200, Australia,
        E-mail: {\tt kuncic@maths.anu.edu.au}},
        E. G. Blackman and M. J. Rees \\ 
        Institute of Astronomy, Madingley Rd, Cambridge CB3 0HA}
\date{}
\pubyear{1996}

\begin{document}

\maketitle

\begin{abstract}
The range of microphysical and global dynamical timescales in the
central regions of Active Galactic Nuclei (AGN) is sufficiently wide
to permit the existence of multiphase structure.
In particular, very dense, cool clouds can coexist with a hot,
magnetically-dominated medium and can thereby efficiently reprocess
the continuum radiation generated in this primary source region.
The strong dynamical forces in this central magnetosphere can give
rise to extremely small clouds.
Microphysical processes then determine whether such clouds can indeed
survive, in spite of their extremely contrasting properties relative
to the surrounding environment, for long enough to produce potentially
observable thermal reprocessing signatures.
We examine specific physical constraints on the thicknesses of such
reprocessing clouds. 
Our results are plotted to show the range of conditions that is
representative of the central regions of AGN.
We find a parameter subspace in the extreme high density regime for
which the effects of microphysical diffusion processes can be overcome
and for which cool gas can maintain pressure equilibrium with the
ambient magnetosphere.
\end{abstract}

\begin{keywords}
galaxies: active -- plasmas, magnetic fields
\end{keywords}

\section{Introduction}

The strong gravitational, radiation and magnetic fields in the central
regions of AGN maintain any gas residing therein in a state of ongoing
dynamical activity.
The relevant microphysical timescales (e.g. thermalization, cooling) can,
however, be extremely short compared to the global dynamical timescale
\cite{Rees84}.
This suggests that an instantaneous `snapshot' could reveal the existence,
albeit for a short time, of small structures close to thermal equilibrium
that are capable of producing reprocessing signatures in the observed
spectra.

Accreting gas exposed to the intense conditions in the central regions
of AGN is unlikely to be homogeneous.
It may consist of cool clumps of material embedded in much hotter and
more tenuous gas (e.g. \citeNP{PustilnikShvartsman74,StellRos84}).
Such a multiphase picture has been successful in accounting for a variety
of thermal reprocessing signatures in AGN spectra (e.g. broad UV emission
and absorption lines, `warm absorber' soft X-ray edges) which are inferred
to arise from gas residing at distances ($.01 - 1$) pc from the central
continuum-forming source (e.g. \citeNP{Davidson72,WTC85,Netzer93}).

The presence of reprocessing material in the form of cool clouds residing
in the central source itself is a particularly interesting possibility
(e.g. \citeNP{Rees87,FerlandRees88,CFR92,Barvainis93}).
The cool clouds are expected to be many orders of magnitude denser than
the reprocessing gas residing further out and the strong confining forces
can give rise to thicknesses which are extremely small by most
astrophysical standards.
Moreover, the clouds may be magnetically confined,
since the primary continuum radiation is expected to be generated in a
hot, dissipative environment that is structured and maintained by 
magnetic fields.
Indeed, the presence of such thin clouds seems to be required in models
for which the field strengths  are at least in
equipartition with the radiation field (e.g. \citeNP{CGF91}).
These unusual physical conditions warrant investigation:
can such small-scale, dense clouds survive in the central magnetospheres
of AGN long enough to be detected through thermal reprocessing signatures
in the observed spectra?

A multiphase medium, with gas at different temperatures and densities
coexisting at roughly the same pressure, can be maintained as long as
the sound-crossing time across the thermal phases is shorter than the
dynamical timescale.
This condition for pressure support imposes an upper limit on the
characteristic thickness of substructure in the form of localized
dense regions cooling radiatively.
The maximum height which can then be attained is just the pressure
scaleheight, as determined by the confining forces of gravity,
radiation and magnetic stresses.
When the dynamical timescale is much longer than the radiative cooling
timescale, gas can maintain thermal and radiative equilibrium and is
thus capable of producing characteristic spectral signatures.
In general, however, microphysical processes threaten the continued
existence of spatially-distinct phases with contrasting physical
properties and therefore impose a stringent constraint on the smallest
thicknesses reprocessing clouds can maintain.
Note that reprocessing clouds require only a very small fraction of the
material in the central engine volume to be compressed, and this material
need only be compressed in one dimension.

In this paper, we determine constraints on the thicknesses of radiative
clouds which can survive against strong dynamical forces and large spatial
gradients in a multiphase environment.
In Section 2, we discuss the constraints on clouds imposed
by their radiative and dynamical state.
In Section 3, we examine mechanisms for cloud confinement and the implied
pressure scaleheight.
We then examine the microphysical diffusion constraints in Section 4.
In Section 5, we present a plot of the cloud parameter space (density
versus thickness) showing regions restricted by the above constraints
for parameters that are representative of the innermost regions of AGN.
We summarize and discuss our results in Section 6.

\section{Radiative and Dynamical Constraints}

Consider a region of thickness $R$, located at a radial distance $r$ from
the central mass in an AGN, in which the gas attains thermal equilibrium,
with a corresponding electron temperature $T_{\rm e}$.
The gas attains an internal thermal and ionization balance when the
timescale $t_{\rm rad}$ for radiative cooling and recombination is much
shorter than the characteristic dynamical timescale, which is of the
order of the timescale $t_{\rm ff}$ for global free-fall when gravity
dominates.
For bremsstrahlung emission, very high electron densities are required
to satisfy $t_{\rm rad} \ll t_{\rm ff}$, with
\[
n_{\rm e} \gg 10^{16}
          \left( \frac{T_{\rm e}}{T_{\rm vir}} \right)^{1/2}
          \left( \frac{r}{r_{\rm g}} \right)^{-2}
           M_{7}^{-1} \mbox{ cm}^{-3} \, ,
\]
where $T_{\rm vir} = m_{\rm p}v_{\rm ff}^{2}/k$ is the virial
temperature corresponding to a free-fall speed
$v_{\rm ff} = c (r/r_{\rm g})^{-1/2}$ and
$r_{\rm g} = GM/c^{2} \! \approx \! (1.5 \times 10^{12})M_{7}$ cm is
the gravitational radius of a central mass $M = 10^{7}M_{7}M_{\odot}$.

Cool, dense clouds embedded in a hot, magnetically-dominated gas pervading
the central continuum-forming region can only be sustained if the
sound-crossing timescale for pressure equilibration with the surrounding
environment is shorter than  $t_{\rm ff}$.
This implies an upper bound on the characteristic thickness $R_{\rm cld}$
of a cloud of gas,
\[
R_{\rm cld} < t_{\rm ff} v_{\rm s} \, ,
\]
where $v_{\rm s}$ is the internal sound-crossing speed at which the
perturbations traverse the structure.
Since gas in the central environments of AGN is most likely to be
magnetized, the total internal sound speed is 
  $v_{\rm s}^{2} = c_{\rm s}^{2} + v_{\rm A}^{2}
                 = (2kT_{\rm e}/m_{\rm p})(1 + 1/\beta)$,
where $c_{\rm s}$ is the thermal sound speed, $v_{\rm A}$ is the
Alfv\'{e}n speed and where $\beta \equiv c_{\rm s}^{2} / v_{\rm A}^{2}$
is the usual plasma beta parameter for the ratio of the thermal to
magnetic pressure inside the gas.

Spatially-localized regions of dense, cool gas can be supported in an
extended atmosphere by dynamical forces other than gravity; these
generally set a more stringent limit to $R_{\rm cld}$ than the
sound-crossing-time constraint mentioned above.
We next discuss some of the dynamical effects in AGN which can provide
confinement mechanisms that enable discrete regions of cool, dense gas
to coexist with an ambient hot, magnetically-dominated medium in a
multiphase system.

\section{Cloud Confinement}

Cloud survival relies on a confinement mechanism to prevent rapid
expansion and dispersion at the internal sound speed.
The region in which the broad emission lines originate has been most
successfully modeled as a two-phase environment, with cool clouds
in photoionization equilibrium embedded in a hot, inter-cloud gas in
Compton equilibrium \cite{KMT81}.
However, thermal pressure alone is insufficent to confine these clouds if
the external medium is heated only by Compton scattering \cite{Fabian86}.
While additional heating mechanisms may be operating
\cite{MathewsFerland87}, the strong radiation and magnetic fields which
regulate the gas dynamics may provide the dominant contribution to the
supporting pressure and confinement for clouds in radiative equilibrium
\cite{Rees87}.
Magnetic fields and radiation pressure also appear to provide a
physically reasonable solution to the problems of confinement and
acceleration of broad absorption line clouds in radio-quiet quasars
\cite{BdKS91}.
Magnetic confinement
is even more likely to be the case for reprocessing clouds in the
central continuum-forming region, where magnetic fields are thought to
provide the chief means of energy dissipation, so that the field
strengths are at least in equipartition with the radiation field
(e.g. \cite{Heyvaerts92}).

\subsection{The Pressure Scaleheight}

The pressure scaleheight for thermal gas in the presence of strong
gravitational, magnetic and radiation fields is deduced from the
standard equation of magnetohydrostatic equilibrium,
\[
n_{\rm e} m_{\rm p} g_{i} \left( 1 - \frac{g^{*}}{g} \right) =
  \nabla_{i} \left( n_{\rm e}kT_{\rm e} + \frac{B^{2}}{8\pi} \right) -
  \nabla_{j} \left( \frac{B_{j}B_{i}}{4\pi} \right)  \, ,
\]
where $g=GM/r^{2}$ is the gravity associated with the central mass
and $g^{*}$ is an effective gravity describing the bulk acceleration
due to internal radiation pressure resulting from photons as they
traverse through and interact with the thermal gas.
We discuss this effect in more detail below.

In the absence of dynamical forces other than gravity, the pressure
scaleheight deduced from the equation of force balance is just
\[
\frac{h}{r} = \frac{v_{\rm s}^{2}}{v_{\rm ff}^{2}}
            = \frac{T_{\rm e}}{T_{\rm vir}}
              \left( 1 + \frac{1}{\beta} \right) \, .
\]
This corresponds to a self-consistent solution for a large-scale,
optically-thin and magnetically-dominated atmosphere, with
$v_{\rm A} \la v_{\rm ff}$.
In the AGN context, such an atmosphere may form above the
inner regions of an accretion disk owing to buoyant poloidal field
lines, which provide an effective means of transferring stresses
from the accretion flow (e.g. \citeNP{Heyvaerts92}).

In addition to hot, diffuse particles, clumps of denser and cooler
material can also be present.
Magnetic pressure can prevent such material from expanding laterally
across the field lines.
Consequently, the material elongates in alignment with the field, forming
narrow filaments.
In regions where the field lines are predominantly open, the filamentary
clouds can be accelerated outward by surface forces in an accompanying
wind or jet (e.g. \citeNP{EmmBlanShlos92}).
In regions where the radial component of the field is small (i.e. tangled
field lines or field loops), clouds can be locally confined by magnetic
stresses due to a discontinuity in the tangential component of the field
in much the same way as prominence structures are thought to survive in
the solar corona (e.g. \citeNP{KippSchlut57}).
As we will discuss below, however, a physically different picture can
arise when a strong radiation field is also present, as is the case for
the central environment of AGN.

Clouds coupled to field lines can maintain a range of possible thermal
pressures whilst maintaining total (thermal + magnetic) pressure
continuity across the boundary,
\[
n_{\rm cld}kT_{\rm cld} \left( 1 + \frac{1}{\beta_{\rm cld}} \right)
  \sim
n_{\rm hot}kT_{\rm hot} \left( 1 + \frac{1}{\beta_{\rm hot}} \right) \, ,
\]
where the subscripts `cld' and `hot' refer to the cool cloud and hot
intercloud phases, respectively.
This condition fixes the degree of magnetic coupling of clouds to plasma
beta values such that
\[
1 + \frac{1}{\beta_{\rm cld}} \sim
    \frac{\tau_{\rm hot}}{\sigmaT n_{\rm cld}}
            \frac{m_{\rm p}g}{kT_{\rm cld}}  \, ,
\]
where $\tau_{\rm hot} \sim \sigmaT n_{\rm hot} h_{\rm hot}$ is
the Thomson scattering optical depth of the external hot plasma.
The observed high-energy spectra imply $\tau_{\rm hot} \sim 1$
(e.g. \citeNP{Zdziarski95}), which is also required for consistency
with Eddington-limited quasi-spherical accretion \cite{Rees84}.

The condition of pressure equilibrium in a multiphase AGN magnetosphere
thus implies a maximum cloud density at which the internal pressure is
entirely thermal (i.e. $\beta_{\rm cld} \gg 1$), with typical parameters
implying
\[
n_{\rm cld} \la \frac{\tau_{\rm hot}}{\sigmaT}
            \frac{m_{\rm p}g}{kT_{\rm cld}}
\]
\[
\hspace{0.7truecm}
   \sim 10^{18} \tau_{\rm hot}
   \left( \frac{T_{\rm cld}}{10^{5}{\rm K}} \right) ^{-1}
   \left( \frac{r}{10 r_{\rm g}} \right)^{-2} M_{7}^{-2}
   \mbox{ cm}^{-3} \, .
\]
Beyond this maximum density at which a cloud becomes essentially a
field-free or `diamagnetic' plasmoid, the internal cloud pressure
exceeds the pressure in the external magnetosphere and hence, confinement
is no longer possible.

\subsubsection{Diamagnetic Effects}

When decoupled from an external magnetic field, clouds distort the
surrounding field lines and subsequently attain a filamentary structure
aligned to the field owing to lateral compression by the restoring magnetic
stresses.
At the same time, the clouds experience a radial force due to large-scale
spatial gradients in the nonuniform field.
This force can effectively eject diamagnetic plasmoids from regions of
high field strength.
The mechanism is often referred to as the `melon-seed effect', since the
combined action of magnetic stresses effectively squeezes out a cloud
between radial field lines like a melon seed between two fingers
\cite{SevKhok53,Schluter57,Parker57}.  This process has been studied in
the context of mass ejections from the solar corona
\cite{Brueckner80,Pneuman83}.

Under the condition of pressure equilibrium between the diamagnetic cloud
and the external, magnetically-dominated medium, the equation of motion
is \cite{Brueckner80,Pneuman83} 
\[
\frac{{\rm d}v}{{\rm d}t} = - g
  \left[ \frac{kT}{m_{\rm p}g}
  \nabla \left( \ln{ \frac{B^{2}}{8\pi} } \right) - 1 \right]  \, .
\]
For a radially-diverging field, with $B \propto r^{-\varrho}$, the
acceleration exceeds gravity when the initial temperature of the
plasmoid is higher than the escape temperature
$T_{\rm esc} / T_{\rm vir} \sim 1/2 \varrho$.
The cool clouds considered here, however, have $T \ll T_{\rm esc}$;
and the melon-seed effect is
unimportant compared with the competing effects of gravity and radiation
pressure.

\subsubsection{Radiation Pressure Effects}

The radiative opacity of dense, cool gas can be sufficiently large
that small-scale clouds subjected to the powerful radiation field in
the central regions of AGN can experience strong volumetric radiation
forces (e.g. \citeNP{deKoolBegel95}).
These forces result from the momentum imparted by photons that are
absorbed and also from the increase in the internal pressure as photons
are subsequently created.
The bulk acceleration in the cloud rest frame is equivalent to an
effective `negative gravity', defined by
\[
\frac{g^{*}}{g} \approx \int {\rm d}\nu \,
  \frac{\sigma_{\nu}}{\sigmaT} \frac{L_{\nu}}{L_{\rm Edd}} \, ,
\]
where $\sigmaT$ and $\sigma_{\nu}$ are the scattering and
radiative opacities, respectively (we note that $\sigmaT$
reduces to the Klein-Nishina relation at high energies).

Since radiation pressure depends on the effective cross-section per
particle, the acceleration is only effective over an optically-thin layer.
Thus, dense clouds with line of sight column densities
$\ll 10^{24}{\rm cm}^{-2}$ for which $\sigma_{\nu} \gg \sigmaT$ can attain
very small scaleheights, with
$h_{\rm cld}$ reduced by a factor $g^{*}/g \gg 1$.
Such clouds can be prevented from being radiatively driven outward by the
tension in field lines.
The compressive action of the radiation force against the restoring
magnetic tension force then gives rise to thin clouds that are highly
favourable for efficient thermal reprocessing of radiation, since a cloud
can span a large covering area whilst at the same time maintaining a small
volume-filling fraction.

As pointed out by \citeN{CFR92}, the opacity of dense, optically-thin
clouds in the central regions of AGN is expected to be dominated by
free-free absorption, which has a  a cross-section
$\sigma_{\nu}^{\rm ff}$ given by
\[
\frac{\sigma_{\nu}^{\rm ff}}{\sigmaT}
  \sim n_{\rm cld} k T_{\rm cld} \frac{r_{0}c^{2}}{h \nu^{3}}
  \left( \frac{k T_{\rm cld}}{m_{\rm e} c^{2}} \right) ^{-3/2}
  \left[ 1 - \exp \left( - \frac{h \nu}{k T_{\rm cld}} \right) \right] \, .
\]
Since the strongest coupling is between the dense gas and low-energy
photons, the strongest radiation pressure effects result from a high
brightness temperature radiation field, such as that generated by a
synchrotron source.
With the presence of strong magnetic fields, the centres of AGN provide
a natural environment for such a nonthermal source and the
synchroton radiation is expected to provide the `seed' field for
subsequent thermal reprocessing by surrounding Comptonizing particles
as well as by an underlying accretion disk if $\tau_{\rm hot} \sim 1$.

\begin{figure}
\epsfysize=0.47\textwidth
\setbox1=\vbox{\epsfbox{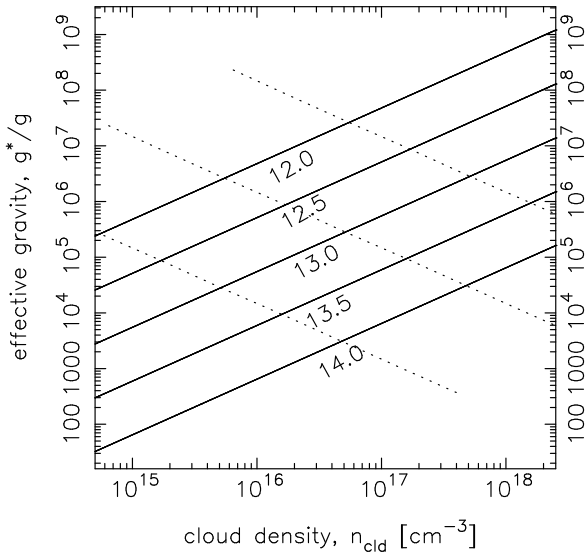}}
\rotr1
\caption{Plot of effective gravity $g^{*}$ (in units of standard gravity g)
due to radiation pressure on clouds with density $n_{\rm cld}$ which
free-free absorb synchrotron radiation.
The logarithm of the self-absorption turnover frequency,
$\log \nu_{\rm t}$, is indicated on each line and the cloud temperature
is set at a constant value $T_{\rm cld} = 10^{5}$ K.
The dotted lines represent maximum cloud thicknesses of $10^{4}$cm, $10^{2}$cm
and 1 cm as $g^{*}/g$ increases.}
\label{fig:effgrav}
\end{figure}

For a self-absorbed synchrotron source, the spectrum is characterized by
a turnover frequency $\nu_{\rm t}$ which invariably lies in the FIR band
\cite{Begelman88}.
The effective gravity due to the free-free opacity can then be
estimated in terms of the luminosity $L_{\rm abs}$ integrated over the
self-absorbed regime of the spectrum of primary radiation which is
reprocessed by the dense clouds.
The free-free absorption opacity evaluated in the Rayleigh-Jeans limit
then implies
\[
\frac{g^{*}}{g} \sim \frac{\sigma_{\nu_{\rm t}}^{\rm ff}}{\sigmaT}
     \frac{L_{\rm abs}}{L_{\rm Edd}} \, ,
\]
which is the factor by which the pressure scaleheight of optically-thin
gas can be reduced from its usual value in the absence of radiation
pressure effects.
Fig. \ref{fig:effgrav} shows how large these factors can become for the
parameters typically expected.
Although small thicknesses can be attained, the clouds are
unlikely to exist as infinitessimally thin structures, since their survival
is then threatened by the microphysical processes which can disrupt
multiphase structure.

\section{Microphysical Constraints}

A multiphase medium in which cool, dense clouds confined by magnetic fields
are immersed in a hot, scattering-dominated plasma cannot be maintained
indefinitely.
The spatial gradients in both temperature and magnetic field trigger
diffusive processes which, despite operating on microphysical scales, can
ultimately result in macroscopic effects (see \citeNP{BegelMcKee90}).
Diffusive mixing acts to restore a multiphase system to a spatially
homogeneous and thermally unstable state.
Over a timescale $\Delta t$, diffusion disrupts the boundary layer
between phases over an effective depth $d \sim \sqrt{D \Delta t}$,
where $D$ is the diffusion coefficient.
Clearly, the most disruptive effects occur when $d$ becomes comparable to
or even exceeds the characteristic thickness $R_{\rm cld}$ of a cloud.
In the present context, it is also particularly important to consider the
possibility that $R_{\rm cld}$ may be smaller than the effective mean free
path for encounters between the hot and cool electrons, in which case a
non-diffusive description of the mixing is required.

\subsection{Thermal Conduction}

The high thermal conductivity of an ambient hot plasma poses the most
serious threat to the continued existence of a spatially-distinct cool
phase in the form of small-scale dense clouds embedded within it.
The cool clouds can be readily penetrated by the hot electrons, so that
the two phases assimilate towards a new thermal equilibrium that is either
hot or cold, depending on whether the clouds evaporate or whether
condensation prevails \cite{BalbusMcKee82,McKeeBegel90}.

\subsubsection{Diffusive vs. Saturated Conduction}

In the diffusion approximation to thermal conduction, the classical
Spitzer conductivity, $\kappa$, defines the thermal diffusion coefficient
by (see \citeNP{CowieMcKee77})
\[
D_{\rm T} = \frac{2}{5} \frac{\kappa}{n_{\rm e}k}
   = \lambdaT \vT 
\]
where $\lambdaT$ is the effective path length between collisions and
$\vT = (kT_{\rm e}/m_{\rm e})^{1/2}$ is the thermal electron velocity.
This formalism is only valid on scales larger than $\lambdaT$; on
smaller scales, a non-diffusive (i.e. collisionless) treatment is
necessary (see \citeNP{McKeeBegel90} and references cited therein).
The observed high-energy spectra of most AGN suggest that Thomson depths
just below unity are compatible with Compton temperatures that can reach
up to mildly relativistic values
(e.g. \citeNP{Johnson94,Zdziarski95,ZdzJohnMag96}).
At such extremely high temperatures, heat-conducting electrons
can freely penetrate small-scale inhomogeneities in the form of cool
clouds embedded in the magnetosphere.
Moreover, the effective path length, $\lambdaT$, for collisions between
the hot electrons and the denser cloud particles is much shorter than
the mean free path, $\lambda_{\rm mfp}$, for collisions between the hot
electrons themselves, with \cite{Spitzer62}
\[
\lambdaT \sim \lambda_{\rm mfp} \frac{n_{\rm hot}}{n_{\rm cld}}
  \sim ( 3 \times 10^{3} )
  \left( \frac{n_{\rm cld}}{10^{18}{\rm cm}^{-3}} \right)^{-1}
  \left( \frac{T_{\rm hot}}{10^{9}{\rm K}} \right)^{2}
  \, {\rm cm} \, .
\]
Hot electrons which penetrate dense clouds on scales larger than
$\lambdaT$ readily lose their energy to collisions.
Consequently, the heated clouds can expand and thereby effectively
evaporate if the density is below the critical density $n_{\rm crit}$
at which the additional heat input can be efficiently radiated away.
Relative to the density of the hot electrons, this critical cloud density
is
\[
\frac{n_{\rm crit}}{n_{\rm hot}} \sim (2 \times 10^{6})
  \left( \frac{T_{\rm cld}}{10^{5}{\rm K}} \right)^{-1/2}
  \left( \frac{T_{\rm hot}}{10^{9}{\rm K}} \right)^{-1/2} \, .
\]
When $n_{\rm cld} \ga n_{\rm crit}$ and radiative cooling dominates
Coulomb heating, the energy deposited into clouds is removed faster than
a high temperature equipartition can be established.
The impenetrating electrons thus remain cool after having lost their
energy and thereby effectively condense into the clouds. 
Hence, the resulting cloud of dense gas maintains a temperature which
remains considerably cooler than the surroundings.

As mentioned earlier, the diffusion approximation to thermal conduction
breaks down on scales smaller than $\lambdaT$, since the conducting
electrons are unimpeded by collisions.
When the hot electrons enter dense clouds with scale heights
$R_{\rm cld} < \lambdaT$, the energy imparted by dynamical friction
effectively produces a uniform heat input throughout the cool gas.
This saturated heat conduction also results in cloud evaporation,
unless $n_{\rm cld} > n_{\rm crit}$.

\subsubsection{The Thermal Diffusion Depth}

For parameters relevant to the innermost central regions of AGN, the depth
to which cool clouds can be penetrated by a diffusive heat flux due to hot
electrons during a timescale $\Delta t$ (in seconds) is
\[
d_{\rm T} \sim (6 \times 10^{6}) \,
  \left( \frac{n_{\rm cld}}{10^{18}{\rm cm}^{-3}} \right)^{-1/2}
  \left( \frac{T_{\rm hot}}{10^{9}{\rm K}} \right) ^{5/4}
  \! ( \Delta t )^{1/2}   \, {\rm cm} \, ,
\]
Even over a timescale as short as a typical radiative cooling time
of the dense gas, with
$t_{\rm rad} \! \sim (2 \times 10^{-5}) n_{18}^{-1}T_{5}^{1/2}$ s for
bremsstrahlung emission (where $n_{18} = n_{\rm cld} / 10^{18}{\rm cm}^{-3}$
and $T_{5} = T_{\rm cld} / 10^{5}{\rm K}$), this diffusion depth is
larger than $\lambdaT$ by a factor
\[
\frac{d_{\rm T}}{\lambdaT} \sim 10
   \left( \frac{T_{\rm hot}}{10^{9}{\rm K}} \right) ^{-3/4}
   \left( \frac{T_{\rm cld}}{10^{5}{\rm K}} \right) ^{1/4}
   \left( \frac{\Delta t}{t_{\rm rad}} \right) ^{1/2} \, .
\]
Thus, the diffusion approximation to thermal conduction is valid over
all relevant timescales, $\Delta t \ga t_{\rm rad}$ for clouds
with thicknesses $R_{\rm cld} > \lambdaT$.
Of these clouds, those with thicknesses $< d_{\rm T}$ are obliterated
unless $n_{\rm cld} > n_{\rm crit}$ or unless they are
decoupled from the field lines to which the hot, conducting electrons
are also tied.

If the cloud density is higher than the minimum, $n_{\rm crit}$, required
for radiative cooling to efficiently remove the extra heat input, then it
is possible that clouds can remain relatively cool whilst threaded by field
lines to which hot electrons are also tied.
When the cloud thickness exceeds $d_{\rm T}$, the conducting electrons
effectively condense as they infuse into the clouds and lose their energy.
At lower densities, however, clouds coupled to the field lines readily
evaporate and are thus unable to produce any observable spectral features
that can be identified with thermal reprocessing by cool gas.
Alternatively, it is possible that clouds are completely decoupled from
the field lines frozen into the external plasma.
Unless the cool, dense gas and the hot, diffuse plasma are coupled to
discrete field lines, such clouds probably manifest themselves as
distinct, self-enclosed structures, possibly with a negligible internal
magnetic field ($\beta_{\rm cld} \gg 1$) if the thermal pressure alone
is sufficient to roughly maintain equilibrium with the external
magnetosphere.

\subsubsection{The Suppression of Thermal Diffusion}

Regardless of how they are coupled to the ambient magnetosphere, dense
clouds are expected to be thermally decoupled from the external conducting
plasma in directions perpendicular to $\bf B$, where the random motion of
hot electrons is strongly impeded by gyration.
Thermal conduction in this transverse direction is attenuated by a factor
$( 1 + \Omega_{\rm e}^{2} / \nu_{\rm coll}^{2} )^{-1}$
\cite{ChaplanCowling52}, where $\Omega_{\rm e} / \nu_{\rm coll}$ is the
ratio of the electron gyrofrequency to the mean collision frequency.

In the central magnetospheres of AGN, field strengths which are
in equipartition with the radiation energy density are typically
$\sim 10^{4}$ G (e.g. \citeNP{Rees84}), which gives
\[
\frac{\nu_{\rm coll}}{\Omega_{\rm e}} \sim (4 \times 10^{-5})
  \, B_{4}^{-1}
  \left( \frac{n_{\rm cld}}{10^{18}{\rm cm}^{-3}} \right)
  \left( \frac{T_{\rm hot}}{10^{9}{\rm K}} \right) ^{-3/2} \, ,
\]
where $B_{4} \! = \! B/10^{4}$ G.
In spite of the extremely high cloud densities, these field strengths
are sufficient to appreciably suppress collisions across the confining
field lines.
Diffusion still occurs over the relevant timescales, since this ratio
implies that roughly $10^{2} T_{\rm cld,5}^{1/2} T_{\rm hot,9}^{-3/2}$
collisions can take place during the radiative lifetime of the dense gas.
However, the corresponding transverse diffusion distance is negligible
compared to $d_{\rm T}$ for diffusion along the field lines owing to
the strongly chanelled motion of the hot electrons along the lines of
force.

Magnetic fields therefore play two crucial roles in a magnetosphere
composed of multiple coexisting phases, providing a confinement mechanism
for spatially-localized regions of cool gas and also effectively insulating
such clouds from the normal transport of thermal energy from the ambient
hot plasma.
As we will explain in the following, however, there is an additional and
more direct means by which heat can be injected into cool clouds in the
central regions of AGN that is independent of the presence of field lines.

\subsubsection{Pair Plasma Effects}

Heat injection by electron-positron ($e^{+}e^{-}$) pairs is a
particularly important consideration for the survival of cool, dense
clouds in the compact environments at the centres of AGN.
This is because once the compactness parameter
$l \equiv \sigmaT L/ m_{\rm e}c^{3} R$
satisfies $f_{\gamma}l>4\pi$, where $f_{\gamma}$ is the fraction of
primary luminosity that is emitted above 1 MeV, the encounters between
$\gamma$-ray photons which produce pairs can occur anywhere -- even
within clouds themselves.
Thus, the pairs can pervade the entire source region, filling up the
magnetosphere in which the primary radiation is generated, and can
equally affect all clouds, regardless of how they are coupled to the
field lines.

Thermal $e^{+}e^{-}$s are expected to constitute an additional component
of the hot plasma phase which maintains Compton equilibrium at roughly
the same temperature as the ambient electron-ion plasma component.
However, even when $L \ll L_{\rm Edd}$, the scattering off the extra
pairs lowers the effective Eddington limit, so that the hot plasma
may be blown outward by the radiation pressure from the Thomson
scattering off ambient photons (see \citeNP{LZR87}).
Because of annihilation, on the other hand, the lifetime of $e^{+}e^{-}$s
can be quite short when $l$ is large, since the timescale is typically
$t_{\rm ann} \sim (r/c) \tau_{\rm pair}^{-1}$, where the Thomson scattering
pair depth is $\tau_{\rm pair} \sim (f_{\gamma}l/4\pi)^{1/2}$ under
steady-state conditions \cite{GFR83}.

The conductive properties of hot, thermal $e^{+}e^{-}$s are essentially
the same as that of electrons in an ordinary hot plasma at the same
Compton temperature, so that pairs created `in situ' can provide a heat
input into cool clouds.
For pairs created inside the clouds, however, the heat injection cannot
adequately be described as necessarily either diffusive or saturated 
conduction,
since a roughly uniform heat input can be provided before the pairs
annihilate.

Energetic charged particles which enter or, in the case of pairs, are
created within a plasma of lower thermal energy slow down due to
long-range Coulomb interactions.
Extensive studies of solar flares have shown that the rate at which
this thermalization occurs is always much faster than the rate at
which positrons annihilate
(e.g. \citeNP{Crannell76} and references cited therein).
If $\Delta E$ ($\sim 1$ MeV) is the total energy lost by pairs due to
collisions with cool, dense cloud particles, then the minimum volume
heating rate is $\sim n_{\rm cld} n_{\rm pair} \sigmaT c \Delta E$,
where $n_{\rm pair}$ is the pair density.
The corresponding minimum cloud density at which this extra heat input
can be balanced by radiative cooling is then
\[
n_{\rm cld} > ( \mbox{ a few } \times 10^{15})
\]
\[
\hspace{1.0truecm} \tau_{\rm pair}
  \frac{\Delta E}{m_{\rm e}c^{2}}
  \left( \frac{T_{\rm cld}}{10^{5}{\rm K}} \right) ^{-1/2}
  \left( \frac{r}{10 r_{\rm g}} \right) ^{-1} M_{7}^{-1}
  \, {\rm cm}^{-3} \, .
\]
At these high densities, clouds in which hot, thermal $e^{+}e^{-}$s are
created can maintain cool temperatures relative to the surroundings by
efficiently radiating away the energy imparted by the pairs.
At lower densities, cloud evaporation is inevitable.
Once thermalized, the cool pairs quickly annihilate, at a rate which
exceeds $\sim n_{\rm cld} \sigmaT c$ by a Coulomb correction
factor that is roughly an order of magnitude at $\sim 10^{5}$ K
(see \citeNP{Crannell76}).

\subsection{Magnetic Diffusion and Kelvin-Helmholtz Effects}

We have shown above that in the magnetospheres of AGN, the motion of hot
electrons is channeled so strongly that their diffusion across the field
lines is essentially negligible, despite the extremely high densities of
cool clouds confined by the field.
It would therefore seem that magnetic diffusion may not necessarily pose
a too serious threat to the survival of such clouds embedded in AGN
magnetospheres.
However, it is possible that non-collisional processes (e.g. turbulence,
local fluctuations) trigger events which can ultimately lead to the
enhanced diffusion of field lines into very dense (and essentially
field-free) clouds.
Of particular relevance to the central regions of AGN, where strong
dynamical effects are important, is the hydromagnetic Kevin-Helmholtz
instability, which we consider here.

Clouds of dense gas moving relative to an external, magnetically-dominated
medium (i.e. with $v_{\rm s,hot} \sim v_{\rm A}$) are susceptible to
disruption by enhanced diffusion due to the Kelvin-Helmholtz (KH)
instability (see \citeNP{AronsLea80} and references cited therein).
This instability is triggered by hydromagnetic turbulence which forms in
a layer between the sheared phases and which fragments the clouds, thereby
effectively increasing the surface area over which microphysical diffusion
operates.
The turbulent mixing of the phases which ensues can then assimilate
the small-scale clouds much more rapidly than collisions.
The instability is particularly destructive to unmagnetized clouds,
since clouds which contain an internal magnetic field (with
$\beta_{\rm cld} > \beta_{\rm hot}$) are further required to undergo some
small-scale reconnection with the diffusing field lines for the phases to
be thoroughly mixed, so that the timescale for disruption by the KH
instability is expected to be longer for magnetized clouds than for
unmagnetized clouds.

The linear growth rate of the KH instability for dense, unmagnetized
clouds moving at a velocity $\bf v_{\rm cld}$ with respect to an external
hot, magnetically-dominated medium is
\[
\gamma_{\rm KH} = \left[
  ( {\bf k . v_{\rm cld}} )^{2} - k_{\parallel}^{2} v_{\rm A}^{2}
                  \right] ^{1/2}
  \left( \frac{n_{\rm hot}}{n_{\rm cld}} \right) ^{1/2}  \, ,
\]
where $k_{\parallel} = k \cos \theta$ is the component of the perturbation
wavevector $\bf k$ that is parallel to the external magnetic field $\bf B$.
The instability can disrupt moving clouds if the characteristic timescale
for the modes to grow, $t_{\rm KH} \sim 1 / \gamma_{\rm KH}$, is shorter
than the cloud sound-crossing response timescale,
$t_{\rm s,cld} \sim R_{\rm cld} / v_{\rm s,cld}$.
Since the most disruptive perturbations are those with $kR_{\rm cld} \ga 1$,
this condition for instability is equivalent to 
\[
\frac{v_{\rm cld}^{2}}{v_{\rm A}^{2}} \ga 1  \, ,
\]
where we have used the pressure equilibrium condition,
$v_{\rm s,cld} \sim v_{\rm A} (n_{\rm hot}/n_{\rm cld})^{1/2}$.

\begin{figure*}
\hspace{0.1truecm}
\epsfysize=0.7\textwidth
\setbox2=\vbox{\epsfbox{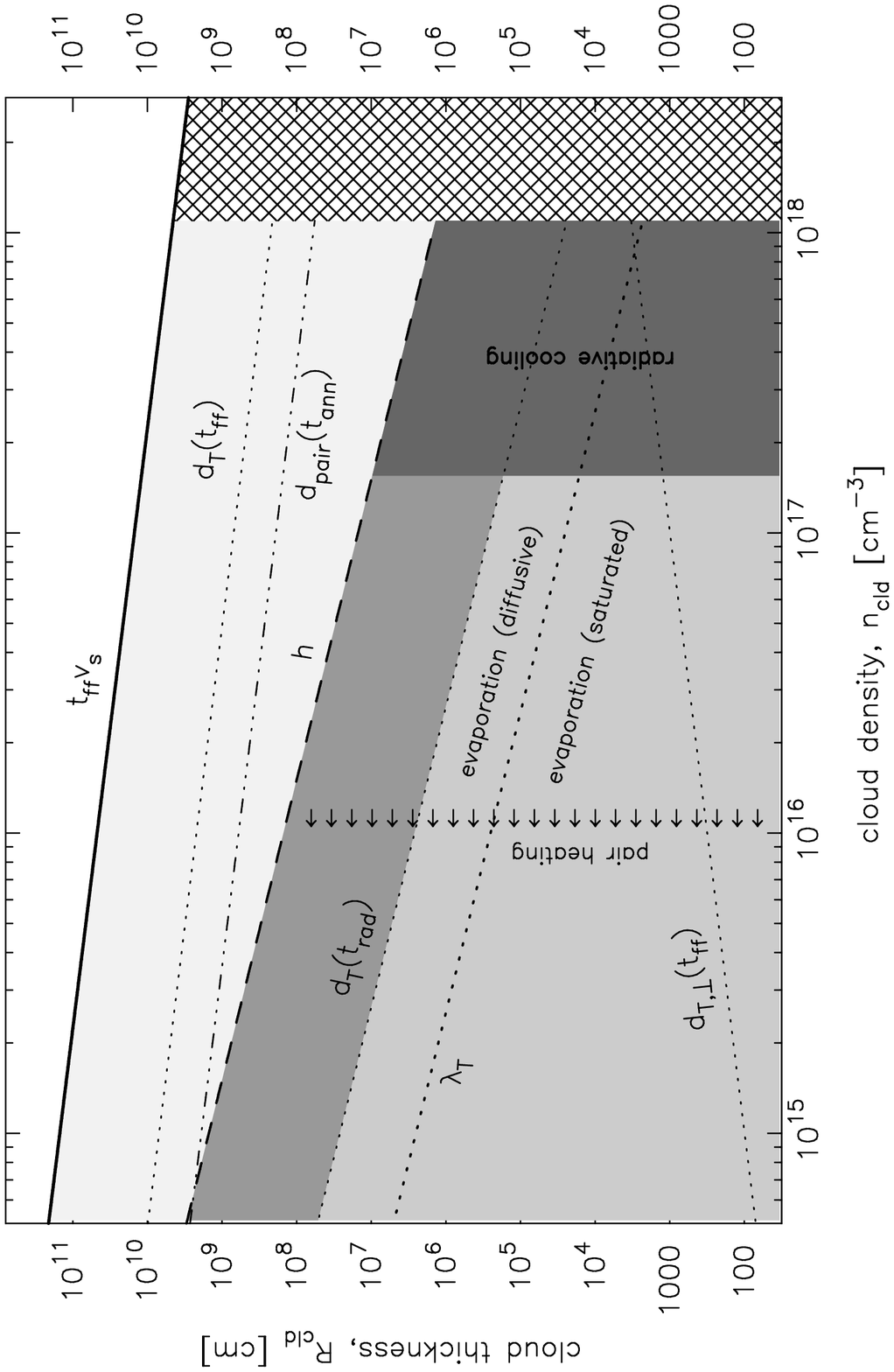}}
\rotr2
\caption{Parameter space for cool, dense clouds in a central AGN
magnetosphere, indicating the densities and thicknesses for which
magnetically-confined clouds can survive in the presence of global
dynamical and microphysical processes.
The cloud temperature is $T_{\rm cld} = 10^{5}$ K and the hot
intercloud gas has a temperature $T_{\rm hot} = 10^{9}$ K and a
Thomson scattering optical depth $\tau_{\rm hot} = 1$.
The compactness parameter of the source region is $l \sim 230$.
The darkest shaded region indicates where cool, dense gas always prevails.
The cross-hatches identify the parameter subspace for which clouds
cannot maintain pressure equilibrium with the external medium.
All labels are defined in the text.}
\label{fig:cfr}
\end{figure*}

Since $v_{\rm A} / v_{\rm ff} \sim h_{\rm hot}/r \la 1$, clouds
accreting towards the central mass in AGN become only marginally
unstable in the absence of dynamical forces other than gravity.
If the clouds experience a net acceleration $g^{*} > g$ due to radiation
pressure force, then
the resulting velocity that can be attained is much larger than
$v_{\rm ff}$ and the clouds become unstable to KH effects.
 
Once the KH instability is triggered, it very rapidly reaches a nonlinear
regime in which the clouds suffer a cascade of fragmentation, with the
enhanced microphysical diffusion processes ultimately sealing their fate.
Because these destructive effects would occur very rapidly, clouds which
are essentially field-free would be rare unless they were being formed
at a much faster rate than clouds coupled to field lines.

\section{The Parameter Space for Cloud Thicknesses}

We now examine how the various physical constraints on cloud thicknesses
determined in the previous sections compare with each other by plotting
the quantities calculated for a range of cloud densities and for a set of
parameters which are represenative of the innermost central regions of AGN.
Under the condition of pressure equilibrium at the boundary between cool,
dense clouds and a hot, magnetically-dominated intercloud plasma, the cloud
plasma beta can be eliminated systematically throughout the above equations
by being replaced with
\[
\beta_{\rm cld} \sim
  \left[ \frac{\tau_{\rm hot}}{\sigmaT n_{\rm cld}}
         \frac{m_{\rm p}g}{kT_{\rm cld}}  -  1 \right] ^{-1} \, ,
\]
where $\tau_{\rm hot} = \sigmaT n_{\rm hot} h_{\rm hot} =
 \sigmaT n_{\rm hot}(kT_{\rm hot}/m_{\rm p}g)(1+1/\beta_{\rm hot})$
is used.
This effectively leaves $n_{\rm cld}$ and $R_{\rm cld}$ as the only
cloud variables once $T_{\rm cld}$ is fixed and the parameters of the
external environment are specified.

Fig. \ref{fig:cfr} shows the resulting parameter space
($n_{\rm cld}$ vs. $R_{\rm cld}$) diagram for cool, dense clouds with a
constant temperature $T_{\rm cld} \! = \! 10^{5}$ K in pressure equilibrium
with a hot, magnetically-dominated intercloud medium with a constant
temperature $T_{\rm hot} \! = \! 10^{9}$ K and with a Thomson scattering
optical depth $\tau_{\rm hot} \! = \! 1$.
The clouds reside in a source region of size
$r \! = \! 10^{13} {\rm cm} \sim 10r_{\rm g}M_{7}^{-1}$ from which emerges a
total bolometric luminosity
$L \! = \! 10^{44} {\rm erg.s}^{-1} \sim 0.1 L_{\rm Edd} M_{7}^{-1}$ and
which thus has a compactness parameter
$l \equiv 4\pi (m_{\rm p}/m_{\rm e})(L/L_{\rm Edd})(r_{\rm g}/r)\sim 230$.
These parameters are representative of current ideas for the central
continuum-forming region of most AGN
(e.g. \citeNP{BlandfordRees92,CFR92,Zdz94,Zdziarski95}).

The parameter space plot ($n_{\rm cld}$ vs. $R_{\rm cld}$) for
cloud thicknesses shows regions that are restricted by the following
relevant constraints: the maximum thickness, $t_{\rm ff}v_{\rm s}$, allowed for
pressure support on a dynamical free-fall timescale (upper heavy line);
the pressure scaleheight, $h$, for confinement by gravity and magnetic
stresses (dashed line); the microphysical limit $d_{\rm T}$ (dotted lines)
due to thermal diffusion calculated for a free-fall timescale, $t_{\rm ff}$,
and for a radiative cooling timescale, $t_{\rm rad}$, as indicated in the
plot; and the effective path length $\lambdaT$ over which collisions are
effective.

The diffusion depth $d_{\rm T,\perp} (t_{\rm ff})$ for thermal
conduction transverse to magnetic field lines calculated for a free-fall
time is also shown.
A direct comparison with $d_{\rm T} (t_{\rm ff})$ for thermal diffusion
along the field lines demonstrates the degree to which the magnetic field
can suppress transverse conductivity for the parameter range of interest.

Also shown is the diffusion depth $d_{\rm pair}$ (dot-dot-dot-dash line)
for thermal pairs along magnetic field lines.
This is the same as $d_{\rm T}$, but calculated for an annihilation
timescale $t_{\rm ann} \sim (r/c)(l/4\pi)^{-1/2}$.
Heat injection by pairs which are created inside clouds is probably
the most serious effect, since all clouds, regardless of their thickness
or their coupling to the ambient magnetosphere, suffer Coulomb heating
before the pairs annihilate unless their density is higher than about
$10^{16}{\rm cm}^{-3}$, when radiative cooling becomes effective.

The darkest shaded region in the plot indicates where radiative cooling
dominates over all other microphysical processes.
Thus, any additional heating due to thermal conduction which takes place
in clouds at these densities can be efficiently radiated away and cool,
dense gas always persists.
thermal conduction leads to cloud evaporation at lower densities and this
is indicated in the plot as either diffusive or saturated (collisionless),
depending on whether $R_{\rm cld}$ is larger or smaller than $\lambdaT$.

The next darkest shaded region in the plot indicates where clouds threaded
by the same field lines to which hot electrons are also tied can survive
the effects of thermal diffusion for timescales $\sim t_{\rm rad}$ which
may be sufficient to produce thermal reprocessing signatures.
Such clouds are unlikely to be able to survive on timescales approaching
$t_{\rm ff}$, since the diffusion depth $d_{\rm T}$ eventually becomes
comparable to the pressure scaleheight and evaporation is unavoidable at
these densities.
Moreover, in the presence of an effective gravity (due to radiation pressure,
for instance), $h$ can be substantially reduced and hot electrons can then
readily diffuse into dense clouds on timescales shorter than $t_{\rm ff}$.
Thus, such clouds can only survive if coupled to seperate field lines or
as self-contained structures, decoupled from the magnetosphere.

\section{Summary and Conclusions}

In light of the mounting observational evidence that thermal reprocessing
plays a crucial role in the central regions of AGN, we have presented
a detailed examination of the possibility that cool, thin clouds coexist
with hot plasma in a central magnetosphere where the primary radiation
is generated.
The thermalization and cooling timescales in these central regions
are sufficiently short compared to the global free-fall timescale
that thermal reprocessing of this primary radiation is inevitable.
Similarly, relatively short sound-crossing times imply that pressure
equilibrium between multiple coexisting phases can be readily
established.
The strong gravitational, magnetic and radiation forces in the central
regions of AGN can then confine spatially-localized regions of gas that
is much denser and cooler than the surrounding, magnetically-dominated
plasma.
The thicknesses that can then be achieved are extremely small by most
astrophysical standards.

The observational relevance of thin clouds with such physical properties
clearly depends on whether they can survive the effects of microphysical
diffusion processes for at least a few radiative timescales during which
they can produce thermal reprocessing signatures in the spectra.
We have summarized the relevant effects in a parameter space diagram
which encompasses the range of cloud densities over which radiative
cooling is important.

Our results indicate that thermal diffusion along the field lines
is the most serious effect, since the diffusion distances
approach the cloud pressure scaleheight on timescales much shorter
than the free-fall time.
This implies that on scales larger than the effective path length between
collisions, clouds that are coupled to the same field lines to which hot
electrons are also tied must be continuously regenerated over timescales
shorter than the free-fall time.

Evaporation is also the fate of most clouds with thicknesses smaller
than the collision path length, since then saturated (non-diffusive)
conduction is effective.
Furthermore, conduction by electron-positron pairs effectively precludes
all clouds with densities below $\sim 10^{16}{\rm cm}^{-3}$, since
radiative cooling cannot then compete with the heating due to encounters
between the pairs and the cloud particles.

Despite the extremely high densities of the clouds, our results show
that the motion of heat-conducting electrons across the lines of forces,
dragging the field lines with them, can still be strongly suppressed.
Cool clouds can thus remain thermally decoupled from the external hot
plasma in directions transverse to the magnetic field.
The diffusion of field lines is then only likely to pose a threat to very
dense clouds which experience a net bulk acceleration due to radiative
or large-scale magnetic forces and which are consequently susceptible
to turbulent diffusion through the Kevin-Helmholtz instability.
Therefore, the clouds most able to reprocess a substantial amount of
radiation and survive against this enhanced diffusion of field lines
are those which are confined by field lines that are sufficiently
tangled to prevent significant bulk motion.

Our results indicate that the most favourable region in the cloud
parameter space lies in the extreme high-density limit, where radiative
cooling dominates all other microphysical process.
This subspace spans a density range $\sim (10^{17}-10^{18})\,{\rm cm}^{-3}$
and a thickness range extending from $\sim 10^{6}$ cm down to microphysical
scales determined by magnetic diffusion.
In this parameter regime, clouds can still be affected by thermal
conduction along the field lines, but the heat input is efficiently
radiated away so that the hot electrons effectively condense into the
clouds and cool, dense gas always prevails.
This gas is therefore capable of producing distinct thermal reprocessing
signatures.

\section*{Acknowledgments}
We thank the referee T. Hartquist for helpful comments.
For financial support, thanks are due to the Cambridge Commonwealth Trust
(ZK), {\small PPARC} (EB) and the Royal Society (MR).

\bibliography{mnrasmnemonic}
\bibliographystyle{mnrasv2}

\end{document}